\documentstyle[epsfig]{mn2e}

\begin{document}

\def\beq{\begin{equation}}
\def\eeq{\end{equation}}
\def\bea{\begin{eqnarray}}
\def\eea{\end{eqnarray}}
\def\dd{\partial}
\def\LL{{\cal L}}
\def\Om{{\Omega_{\rm m}}}
\def\Ob{{\Omega_{\rm b}}}
\def\om{{\omega_{\rm m}}}
\def\equ{{\mathrm equ}}
\def\zt{{z_{\rm t}}}

\title[Optimizing BAO surveys II]{Optimizing
  baryon acoustic oscillation surveys II: curvature, redshifts, and
  external datasets} 

\author[Parkinson et al.]{\parbox[t]{\textwidth}{ David
    Parkinson\thanks{drp21@sussex.ac.uk}$^1$, Martin Kunz,$^1$ Andrew
    R. Liddle$^1$, Bruce A. Bassett$^{2,3}$,  Robert C. Nichol$^4$
    and Mihran Vardanyan$^5$} 
    \\ \\ $^1$ Astronomy Centre, 
  University of Sussex, Brighton, BN1 9QH, U.K. \\  $^2$ South
  African Astronomical Observatory, P.O.\ Box 9, Observatory 7935,
  Cape Town, South Africa\\ $^3$ Department of Maths and Applied
  Maths, University of Cape Town, South Africa\\ 
  $^4$ Institute of Cosmology \& Gravitation,
  University of Portsmouth, Portsmouth, P01 2EG, U.K. \\
$^5$
    Astrophysics Department, 
   University of Oxford, 
  Denys Wilkinson Building, Keble Road, Oxford, OX1 3RH, U.K. }

\maketitle 

\begin{abstract}
We extend our study of the optimization of large baryon acoustic
oscillation (BAO) surveys to return the best constraints on the dark
energy, building on Paper~I of this series (Parkinson et al.\
2007). The survey galaxies are assumed to be pre-selected active,
star-forming galaxies observed by their line emission with a constant
number density across the redshift bin. Star-forming galaxies
have a redshift desert in the region $1.6 < z < 2$, and so this redshift
range was excluded from the analysis. We use the Seo \& Eisenstein (2007) 
fitting formula for the accuracies of the BAO measurements, using only the
information for the oscillatory part of the power spectrum as distance
and expansion rate rulers. We go beyond our earlier
analysis by examining the effect of including curvature on the optimal
survey configuration and  updating the
expected `prior' constraints from Planck and SDSS.  We once again find that the
optimal survey strategy involves minimizing the exposure time and
maximizing the survey area (within the instrumental constraints), and
that all time should be spent observing in the low-redshift range
($z<1.6$) rather than beyond the redshift desert, $z>2$. We find that when assuming a flat
universe the optimal survey makes measurements in the redshift range
$0.1 < z <0.7$, but that including curvature as a nuisance parameter
requires us to push the maximum redshift to 1.35, to remove the
degeneracy between curvature and evolving dark energy.  The inclusion
of expected other data sets (such as WiggleZ, BOSS and a stage III
SN-Ia survey) removes the necessity of measurements below redshift
0.9, and pushes the maximum redshift up to 1.5.  We discuss
considerations in determining the best survey strategy in light of
uncertainty in the true underlying cosmological model.
\end{abstract}
\begin{keywords}
cosmological parameters -- large-scale structure of universe -- surveys
\end{keywords}

\section{Introduction}

The discovery of the accelerating universe, driven by some mysterious
dark energy, has motivated the conceptualization and design of a
number of future surveys that will seek to discover its nature. These
include, but are not limited to: Wide-field Fiber-fed
Multi-Object Spectrograph (WFMOS), the Dark Energy Survey (DES), 
Panoramic Survey Telescope \& Rapid Response System (Pan-STARRS), 
Baryon Oscillation Spectroscopic Survey (BOSS), Large Sky Area Multi-Object
Fibre Spectroscopic Telescope (LAMOST), Hubble Sphere Hydrogen Survey 
(HSHS), Square Kilometre Array (SKA), Large Synoptic Survey Telescope (LSST),
Euclid and the Joint Dark Energy Mission (JDEM). These will
deploy an array of methods to probe the dark energy, such as baryon
acoustic oscillations (BAO), weak lensing, cluster number counts, and  type-Ia
supernovae (SN-Ia).

In such a crowded marketplace it is important to have a compelling
product by demonstrating effective use of resources. In previous
papers some of the present authors have examined the application of
design principles to the construction of new surveys, by optimizing the
surveys to give the best science return (Bassett 2005; Bassett,
Parkinson \& Nichol 2005a).  Also, recently a team commissioned by the
US NSF (the Dark Energy Task Force or DETF) laid out a `roadmap' of
how dark energy experiments may develop into the future (Albrecht et
al.\ 2006), and similar studies have been undertaken by UK and
European funding agencies.

This paper is a continuation of our previous work (Parkinson et al.\
2007, hereafter P07), in which we considered optimizing a BAO survey
similar to the WFMOS design.\footnote{As of May 2009 the original WFMOS project has been terminated through lack of sufficient available funding via the Gemini Observatory, but our methodology and qualitative conclusions are generally applicable to any similar future proposals.} The conceptual design for the WFMOS dark
energy survey is to conduct a large area survey of the sky, measuring
the redshifts of order millions of galaxies. The power spectrum of
these galaxies traces the power spectrum of the underlying matter
density, and this contains the imprint of the primordial sound waves
in the photon--baryon plasma (the BAO). These `wiggles' in the power
spectrum can be used as standard rulers to measure the
angular-diameter distance ($d_{\rm A}$) from those modes transverse to
the line of sight, and the Hubble rate at that redshift ($H(z)$) from
the radial modes. For a description of WFMOS, see Bassett, Nichol \&
Eisenstein (2005b). Very similar surveys have been proposed for LAMOST
(Wang et al. 2009) and the 4-m Mayall telescope (BigBOSS; Schlegel et
al. 2009b).

In P07 we set out the basics of our optimization methodology, defining
the concepts of a Figure of Merit (FoM), very similar to the one
proposed by the DETF but now as a function of survey strategy, and a
survey parameter space, where a particular survey configuration is
defined in terms of time, area and redshift. When these survey
parameters are combined with information about the instrument, we can
predict the number of galaxies that will be observed, the accuracy
with which the BAO will be measured, and the resulting FoM. By
plotting how the FoM varies with survey parameters, we can find the
optimal survey. We found that the optimization preferred the surveys
to be as large in area as possible, limiting the exposure time to be
as small as possible, to beat down the shot noise from limited galaxy
numbers.

In this paper we address the issues of the survey redshift ranges in
the high and low-redshift regimes, and the time spent observing in
each of them. We see how these survey parameters are affected by the
cosmological parameters being considered, and by the other surveys
that are included as priors in the analysis.

In Section \ref{whatwedidlastsummer} we briefly review the details of
our previous optimization, before describing the details of how our
analysis has been updated. In Section \ref{sec:results} we state the
optimal configurations for a WFMOS-like survey by itself.  In Section
\ref{sec:otherdata} we look at the effect on the optimal survey design
of adding in other experimental data as `prior' measurements. In
Section \ref{sec:conclusions} we outline our conclusions.

\section{Optimization Procedure}
\label{whatwedidlastsummer}

\subsection{Survey definition}

\begin{figure}
\center
\epsfig{file=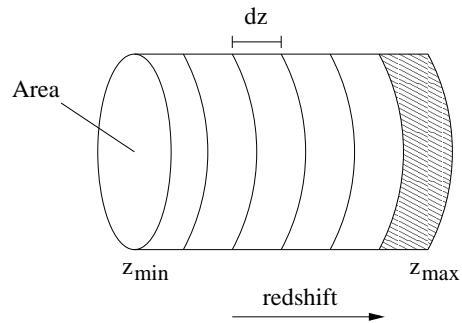, width=6 cm}
\caption[fig:surveyschematic]{\label{fig:surveyschematic} A schematic
 illustrating  how the survey parameters are defined. The survey
 volume is defined by the area on the sky and the minimum and maximum
 redshifts. The redshift range 
 is subdivided into a number of slices
 of fixed width $dz$ for computing the FoM. The number density is fixed by
 the density in the final redshift slice (the shaded region) for a given exposure time.}
\end{figure}

We perform our optimization as described in P07. We consider a set of
allowed survey geometries, described by the parameters listed in Table
\ref{surveyparameters}, and illustrated by a schematic in Figure 
\ref{fig:surveyschematic}.  A general survey is divided into low- and
high-redshift components, the former corresponding to $z<$ 1.6
and the latter $z>$ 2, separated by the redshift desert within
which ground-based surveys cannot effectively obtain redshifts due to
the lack of galaxies emitting in the optical wavelengths. The terminology `low' and `high' has this meaning
throughout.  

\begin{table}
\centering
\caption{List of survey parameters in each redshift regime.
See Parkinson et al.~(2007) for detailed explanations. Note that
we no longer vary the number of redshift bins, but instead
divide up the redshift ranges into thin slices for the FoM calculation.}
\label{surveyparameters}
\begin{tabular}{cc} \hline
Survey Parameter & Symbol \\ \hline
Survey time & $\tau_{\rm low}$, $\tau_{\rm high}$ \\
Area covered  & $A_{\rm low}$, $A_{\rm high}$ \\
Minimum of  redshift bin & $z_{\rm low}$ (min), $z_{\rm high}$ (min)  \\
Maximum of redshift bin &$z_{\rm low}$ (max), $z_{\rm high}$ (max) \\
Number of pointings & $ n_{\rm p}(\rm{low}) $, $n_{\rm p}({\rm high})$ \\  \hline
\end{tabular}
\end{table}

The survey parameters are limited by some constraints, listed in
Table~\ref{constraintparameters}.  These are the same as in P07, with
the exception of the limits on the redshift bins ($z_{\rm min}$,
$z_{\rm max}$), which have been relaxed as we now include a more
reasonable model of the efficiency/response of the WFMOS spectrograph
to light at different wavelengths. The details of this wavelength
throughput are not public, but can be taken to be very similar to that of
the Sloan Digital Sky Survey (SDSS) spectrographs.\footnote{Details of
the SDSS spectrographs can be found at
http://www.astro.princeton.edu/PBOOK/spectro/spectro.htm}

 \begin{table}
\centering
\caption{List of constraint parameters.}
\label{constraintparameters}
\begin{tabular}{cc} \hline
Constraint Parameter & Value \\ \hline
Total observing time & 1500 hours \\
Field of view & 1.5$^o$ diameter \\
$n_{\rm fibres}$ & 3000 \\
Aperture & 8m \\
Fibre diameter & 1 arcsec \\
Overhead time between exposures & 10 mins \\
Minimum exposure time & 15 mins \\
Maximum exposure time & 10 hours \\
Wavelength response &   Priv. comm with AAO   \\ 
Width of redshift slices, $dz$& 0.05 \\ \hline
\end{tabular}
\end{table}

Having established the details of the survey, we compute the total
number of galaxies that will be measured. We assume a pre-existing
source catalogue of photometrically selected galaxies, from which we
can effectively target either line-emission active star-forming
galaxies or passively-evolving continuum emission galaxies. These
details have not changed from the previous paper. Since we found in
P07 that the active star-forming galaxies, whose redshifts are to be
obtained by measuring the O[II] emission line doublet at low redshift
and Lyman-$\alpha$ at high redshift, are the preferred targets, we
adopt these as standard for all the analysis in this paper. We also
set a somewhat arbitrary lower limit of 15 minutes for the exposure
time, representing a reasonable compromise when taking into account
a rather pessimistic estimate of a 10 minute overhead time between
exposures.  We further assume that the galaxies are targeted so as to
generate a sample of uniform number density across each redshift bin.

We also include an estimate of the bias of these galaxies, and its evolution
with redshift. At low redshift we take Wake et al.\ (2008) as our guide, 
assuming the bias (weakly) tracks the linear growth function, using the following
formula
\beq
b(z_{lo}) = 1 + (b(z_{hi}-1)D(z_{hi})/D(z_{lo})\,,
\eeq
where $D(z0$ is the growth function. Here we take $z_{hi}=0.55$ and
$b(z_{hi})=1.3$
  At high redshift
we use the result of Myers et al.\ (2007) that  the bias grows as $(1+z)^2$.

Once the redshift ranges and number of galaxies have been determined,
the cosmological parameter analysis can proceed. Here we slice the
redshift bins into a number of sub-bins, where the width of these
sub-bins is fixed and the number is determined by the redshift range
(as shown in Figure \ref{fig:surveyschematic}).
We take the width of the redshift slices to be constant, $dz=0.05$,
with the redshift range always being an integer number of these slices
and the minimum and maximum redshifts discretized in the same units.

In computing the BAO errors on each slice, we do not include the
possible correlations between slices that may be caused by large-scale
modes in the power spectrum. Our slice width $dz$ is chosen to be fairly
wide to reduce such correlations. These will have the effect of
decreasing the constraining power of the survey and so lowering the
FoM. We do not necessarily expect including these effects to change
the optimal survey, as they will not change the redshifts at which the
measurements are being made, only the accuracy of the measurements.

\subsection{Figure of Merit (FoM)}

Once the area, redshift range and slices, and galaxy number of the
survey have been determined, we can use fitting formulae to estimate
how well the BAO will be measured, and what distance information
will be returned. In Rassat et al.\ (2008) a comparison was made 
between different methods for extracting information from a future
galaxy survey. Here, following on from P07, we only use the
oscillatory part of the power spectrum (the 'wiggles'), as we 
consider this the most robust source of distance information
that can be extracted. The full power spectrum is 
degenerate with primordial power spectrum parameters (tilt, running) 
and also details of the growth of structure on large scales (nonlinear 
bias, non-linear growth). The anisotropy of the power spectrum can 
be used as an Alcock--Paczynski (AP) test, but this require details of 
the non-linear behavior of the redshift-space distortions.

In P07 we used the formula published by
Blake et al.\ (2006), but this has been superseded by those of Seo \&
Eisenstein (2007). We use the formula derived from a Fisher matrix 
approach, using a 2-D approximation of only the oscillatory part 
of the power spectrum (equation 26 in their paper). 
These fitting formula estimate the errors in the position of the baryonic
features along and across the line of sight, as well as the correlation 
between them. They also have the added advantage that they
can simulate the effect of `reconstruction' of the linear oscillations
in the non-linear regime (though we do not use reconstruction in this
paper). This can lead to increased accuracy at lower redshifts, where
non-linear effects on the power spectra are present at the same scales
as the BAO. The accuracies of the BAO measurements leads to the
calculation of the FoM.

In P07, as in the DETF report, the CPL parameterization (Chevallier \&
Polarski 2001; Linder 2003) of the dark energy equation of state 
was used, given by 
\beq 
w(a) = w_0 + w_a(1-a) \,,
\eeq 
where $w_0$ and $w_a$ are adjustable constants.
The FoM we used in the previous paper was the D-optimal
criterion, the inverse of the determinant of the ($w_0$, $w_a$)
covariance matrix, 
i.e.\
\beq 
{\rm FoM_{old}} = \mbox{det}^{-1}({\bf C}) = \frac{1}{\sigma^2_{w_0
w_0}\sigma^2_{w_a w_a} - \sigma^4_{w_0 w_a} }
\eeq 
Here we have switched to the square root of the inverse of the
determinant, bringing us into line with the DETF FoM, 
\beq
{\rm FoM_{\rm new}} = \frac{1}{(\sigma_{w_a}\sigma_{w_{\rm p}})} =
\frac{1}{\sqrt{\sigma^2_{w_0 w_0}\sigma^2_{w_a w_a} - \sigma^4_{w_0
w_a}}}\,,
\eeq 
where $w_{\rm p}$ is the equation of state at the `pivot' redshift.
Hence our new FoM is the square root of our old FoM. We use this new
definition of the FoM throughout.

The FoM is computed using a Fisher matrix
approach. Details are laid out in Appendix \ref{Fisher:appendix}.

\subsection{Adding curvature}

\begin{table}
\centering
\caption{The fiducial cosmological parameters used in this paper.}
\label{cosmologicalparameters}
\begin{tabular}{cc} \hline
Parameter & Value \\ \hline
$w_0$ & $-1$ \\
$w_a$ & 0 \\
$\Omega_{\rm m}$ & 0.3 \\
$\Omega_{\rm r}$ & $8.2\times10^{-5}$\\
$\Omega_k$ & 0 \\
$\Omega_{\rm DE}$ & $1-\Omega_k-\Omega_{\rm m}-\Omega_{\rm r} $ \\
 $H_0$ & 70 \\ 
 $\Ob$ & 0.0441 \\
 $n_{\rm s}$ & 1 \\
 $\sigma_8$ & 0.9 \\ \hline
\end{tabular}
\end{table}

We have expanded our cosmological parameter space from P07, by
including the effect of curvature on our analysis. The importance of
doing so has been emphasized by Clarkson, Cortes \& Bassett 
(2007), who showed
than even a small curvature can seriously bias dark energy measurements.
Our cosmological
parameter space ($\Theta$) is now defined to be 
\beq \Theta = \{ w_0,
w_a, \Omega_{\rm DE}, \Omega_k, h, \Ob h^2, n_{\rm s}\} \,.  
\eeq 
The fiducial values for these parameters are given in
Table~\ref{cosmologicalparameters}. Additional parameters not allowed
to vary are the radiation energy density $\Omega_{\rm r}$ and the
matter spectrum normalization $\sigma_8$. Note that the `wiggles-only'
method of BAO does not constrain $\sigma_8$ directly, it is included
here as it is an input parameter into the Seo \& Eisenstein (2007) fitting formula.

We include the measurements of the BAO by SDSS and 2dF (Eisenstein et al.\ 
2005, Percival et al.\ 2007) as prior information.  The SDSS/2dF
prior is the ratio of measurements of $D_{\rm V}$ at $z=0.2$
and $z=0.35$, where $D_V$ is defined as
\beq
D_V \equiv [r^2cz/H(z)]^{1/3} \,.
\eeq
The accuracy of this measurement is given in Percival et al.\ (2007).

We also include measurements of the CMB by the Planck
satellite (Mukherjee et al.\ 2008) as prior information. Planck 
will make accurate  measurements of the distance to last scattering ($R$) 
and the position of the first peak ($l_a$),  defined to be
\beq
R \equiv \sqrt{\Omega_{\rm m} H_0^2}\, r(z_{\rm CMB}) \,,~~~l_a \equiv
\frac{\pi r(z_{\rm CMB})}{r_s(z_{\rm CMB})} \,. 
\eeq
The accuracy of these measurements was estimated by simulating
temperature and polarisation power spectra (i.e. using TT, TE and EE), 
and running Markov Chain Monte Carlo
(MCMC) chains
to estimate the error on $R$ and $l_a$, as described in Mukherjee et al.\ 
(2008). The results of this paper requires us to include $\Ob h^2$ and $n_s$ in our
analysis.\footnote{We could have included a prior from Big Bang
Nucleosynthesis, which yields $\Ob h^2
= 0.0214 \pm 0.0020$. However, Planck will constrain $\Ob h^2$ well
enough that such a prior has a minuscule effect on the FoM, changing
only the 4$^{th}$ decimal place.} We model the predicted Planck likelihood
using the covariance matrix on these four parameters ($R$, $l_a$, $\Ob
h^2$ and $n_s$) given in Mukherjee et al.\ (2008). Note that
the constraints on $\Ob h^2$ and $n_s$ only come from the CMB,
and these parameters would not be constrained by BAOs only.
There is a loss of information in considering only the constraints on
these four parameters rather than the full CMB power spectra, but 
Mukherjee et al.\ (2008) found that when considering constraints
on Dark Energy models and combining this condensed form of
information with other distance probes this information loss 
is negligible.

\subsection{Searching the parameter space}

We search through the survey parameter space as in P07 using 
simulated annealing, with long
MCMC chains undergoing thermodynamic
scheduling (see Cerny 1985) to push them closer to the optimum. The
FoM takes on the role of the likelihood in parameter estimation MCMC,
where the probability of moving from a survey configuration ($s$) to a
new one ($s'$) is given by
\beq
P(s \rightarrow s') = {\rm min}\left\{1, \frac{{\rm FoM}(s')}{{\rm
    FoM}(s)}\right\}\,. 
\eeq
By employing thermodynamic scheduling, we introduce a temperature
$T$ that modulates the probability of acceptance, thus 
\beq
P(s \rightarrow s') = {\rm min}\left\{1, \left(\frac{{\rm
    FoM}(s')}{{\rm FoM}(s)}\right)^{1/T}\right\}\,. 
\eeq
As the temperature of the chain goes from `hot' to `cold' the
probability of accepting a survey with a smaller FoM rapidly
diminishes. This technique is employed for example in Wit, Nobile \&
Khanin (2005), and has also been used in optimizing cluster
surveys for probing the Dark Energy in Wu, Rozo \& Wechsler (2009),
which appeared after this paper.

The nature of optimization is that we are interested only in a tiny
region of the survey parameter space, and so large numbers of chain
elements are not necessarily a guarantee of reaching the true global
optimum. Some of the parameters we have included may have only small
contributions to the FoM, or may actually be detrimental (e.g. observing at
high redshift may reduce the FoM as it reduces the time that can be spent observing
at low redshift). We therefore often run refinement searches, in a 
lower-dimensional parameter space, to speed up 
reaching the optimum. When the
MCMC search indicates that some of the parameters (time spent at high
redshift, exposure time etc) can be set to specific values, a follow-up
search is carried out with these parameters fixed to refine the
optimal survey.

One aim of the optimization is to discover how far we can push the
survey away from the optimum configuration without degrading the
performance too much. For this we introduce `flexibility bounds',
which describe or delimit the region of parameter space where the FoM
has fallen to 90\% and 60\% of the optimum value. This idea
was introduced in P07, but as we have changed our definition of FoM
from that paper, we have also changed our definition of the
flexibility bounds in line with that. The flexibility bounds are interesting as they
show the relationship between  survey parameters, such as the
survey area and time.

\subsection{Effect of methodology changes from Paper~I}

As compared to P07, this paper makes significant changes to the
methodology. On the observational side is an improved understanding of
the WFMOS instrument, the adoption of the BAO fitting formulae of Seo
\& Eisenstein (2007), and the improvement of prior information from
SDSS and expected from Planck. On the theoretical side is the
inclusion of curvature within the cosmological model.  To illustrate
the effect of these changes, we consider a `standard', non-optimized,
WFMOS survey outlined in Table \ref{standard:table}. This is intended
to represent the sort of survey assumptions one might have made
without optimizing. We compare three different calculations of the
FoM.

\begin{table}
\centering
\caption{Survey parameters and FoMs for a `standard' (non-optimized) survey (including Planck and SDSS as prior information).} 
\label{standard:table}
\begin{tabular}{lc} \hline
Survey Parameter & Value \\ \hline
$A_{\rm low}$ (sq. degrees) & 2000  \\
$\tau_{\rm low}$ (hours) & 800 \\
$z_{\rm low}$ (min) & 0.5\\
$z_{\rm low}$ (max)&   1.3 \\
exposure time (minutes) & 32 \\
number density $({\rm Mpc}^{-3} h^3) $ & $8.3\times 10^{-4}$ \\
number of galaxies  & $3.4\times10^6$ \\
$A_{\rm high}$ (sq. degrees) & 300 \\
 $\tau_{\rm high}$ (hours)& 700\\
$z_{\rm high}$ (min)& 2.3\\
  $z_{\rm high}$ (max) & 3.3\\
exposure time (minutes) & 237  \\
number density $({\rm Mpc}^{-3} h^3) $ & $4.5\times 10^{-4}$ \\ 
number of galaxies  & $5.5 \times 10^4$ \\\hline
FoM (old method, flat Universe) & 7 \\
FoM (new method, flat Universe) & 18\\ 
FoM (new method, with curvature) & 9 \\ \hline
  \end{tabular}
  \end{table}

Under the old methodology, with a flat Universe, the FoM was
7. Improved understanding of the instrument and additional
prior information has indicated that it will be significantly more
powerful; under the same flat Universe assumption the FoM is improved
to 18, substantially reducing the uncertainty of each of the two dark
energy parameters. Inclusion of curvature, however, necessarily
degrades the outcome, lowering the FoM to 9.

\section{WFMOS optimal surveys}
\label{sec:results}

We break the results of the analysis down into the following
subsections. In Section \ref{ssec:previous} we review the results from
our previous work, then in Section \ref{ssec:curvature} we look at the
effect of adding curvature on the time split between the two redshift
regimes. In Section \ref{ssec:strategy} we discuss the best survey
strategy in light of uncertainty in the true underlying cosmological
model.  In this section we only assume priors from SDSS
and Planck.

It would be possible to do an optimization for the experiment
without any prior information, but it would be misleading 
to carry out  such an optimization. The 
SDSS data already exists, and the Planck data will do soon 
(and even WMAP measurements might be good enough for 
this purpose) and the principal goal of optimization is to find the 
correct niche for an experiment. If a survey such as WFMOS were 
forced to spend time observing at high redshift to remove the 
degeneracy with curvature, when it could just as easily do so by 
incorporating the results from Planck, this would be waste of time 
and resources, and the incorrect kind of optimization to perform. 
It is therefore imperative to account for all relevant 
information already available when optimizing a future survey.

\subsection{Previous work}
\label{ssec:previous}

In our previous paper we conducted an analysis where we varied only
four parameters ($w_0$, $w_a$, $\Om$ and $\Om h^2$), assuming the
Universe to be flat so the dark energy density is given by
$\Omega_{\rm DE} = 1-\Om$. We found that the optimization preferred to
concentrate all the survey time into the low-redshift regime. We also
found that short exposure times of just a few minutes on an 8-metre
class telescope are sufficient to obtain redshifts for the majority of
line-emission galaxies. Although a longer exposure time would result
in higher quality spectra with fewer wasted fibres, it also reduces
the amount of area that can be surveyed during a fixed total survey
duration. A large area is important to maximize the number of surveyed
galaxies, so that the shot noise can be beaten down.  The optimal
surveys were therefore driven to the smallest allowed exposure time
of 15 minutes. The best strategy was to ignore the possibility of
high-redshift observations, as there were no parameter degeneracies
that required such observations to break them.

\begin{table}
\centering
\caption{Optimal survey parameters, optimized for a flat universe and including curvature as a nuisance parameter. The FoM is computed including prior information from Planck and SDSS, as described in the text.
The parameters for the high redshift bin are not included as the optimal surveys spend all their time observing at low redshift. We also include the one sigma errors on the cosmological parameters predicted by the Fisher matrix approach for these optimal surveys.} 
\label{table:flatcurved}
\begin{tabular}{lcc} \hline
Survey Parameter & Flat & Curved  \\ \hline
${\rm A}_{\rm low}$ (sq. degs) & 6300 & 6300   \\
$\tau_{\rm low}$ (hours) & 1500 & 1500 \\
$z_{\rm low}$ (min) & 0.1 & 0.1 \\
$z_{\rm low}$ (max) & 0.7 & 1.35\\
exposure time (mins) &15.0  & 15.0 \\
number density  ($h^3$/Mpc$^{3}$) &$3\times10^{-3}$ & $6.6 \times 10^{-4}$ \\ 
number of galaxies  &$10.8 \times 10^6$  & $10.8 \times 10^6$ \\ \hline
FoM & 57  & 32   \\  \hline
$\sigma (w_0)$ & 0.14 & 0.23\\
$\sigma (w_a)$ & 0.44 & 0.70 \\
$\sigma (\Omega_{\rm DE})$ & 0.012 & 0.018 \\
$\sigma (\Omega_{k})$ & - & $2.5 \times 10^{-3}$ \\ \hline

\end{tabular}
\end{table}

We reanalyzed this case, taking into account all the
improvements we had made, but still assuming a flat
universe where the curvature is not included as a nuisance parameter
in the calculation of the FoM. We recovered very similar
results to those given in P07. The optimal survey parameters
are given in Table~\ref{table:flatcurved}, and the FoM as a function of the
survey parameters (for the low-redshift bin only) is shown in
Figure \ref{flat_hilo_1d}. The detailed results of this analysis are shown in Figure
\ref{flat_hilo} in Appendix~B.

\begin{figure}
\center
\epsfig{file=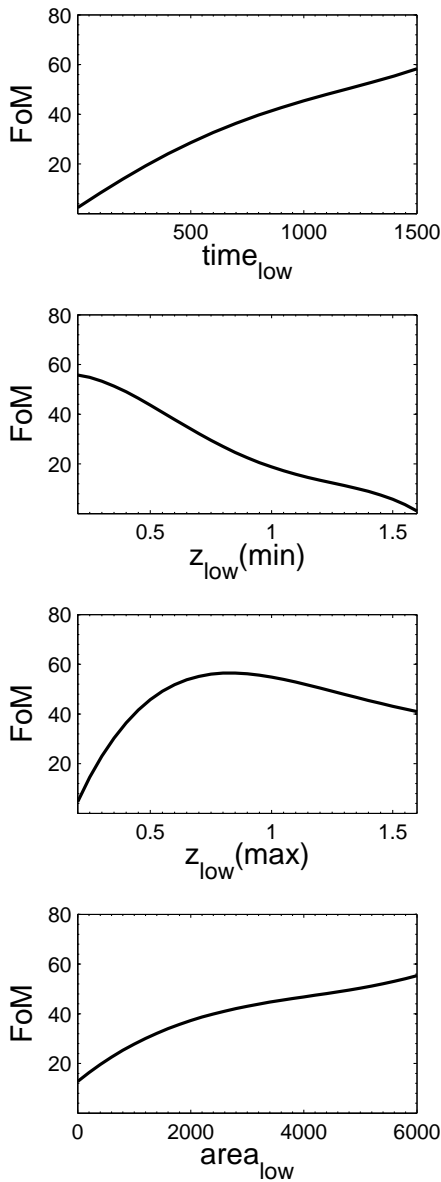, width=5.9 cm}
\caption[flat_hilo_1d]{\label{flat_hilo_1d} The FoM as a function of
  the survey parameters, where the surveys have been optimized
  assuming a flat universe. The other parameters were chosen to
  maximize the FoM. The area and redshifts of the high-redshift bin
  have been omitted, as the optimal survey spends all its time
  observing at low-redshift.}
\end{figure}

The optimized survey represented a
substantial gain in FoM with respect to the `standard' survey.

\subsection{Adding curvature as a parameter}
\label{ssec:curvature}

We now want to understand the effect of including the curvature of the
Universe $\Omega_k$ as a free parameter in our analysis on the best
survey. We continue to assume a flat fiducial cosmology, but now
require our observations to also constrain curvature.

In Table \ref{table:flatcurved} we show the optimal survey allowing for
curvature as a nuisance parameter.  We see that the optimal FoM is
reduced, as the inclusion of an extra parameter ($\Omega_k$) degrades
the constraints on $w_0,w_a$.  We find once again that the preferred
survey is one that spends all its time observing at low redshift.
However, in contrast to the flat case, we see that the maximum
redshift of the low-redshift bin is forced up to $z=1.35$.  BAO
measurements at these higher redshifts are needed to remove the
degeneracy between evolving dark energy and curvature.

The FoM as a function of the survey parameters is shown in
Figure~\ref{curved_hilo_1d}, and the detailed results of this analysis are shown in Figure
\ref{curved_hilo} in Appendix~B. The position of the minimum of the
redshift bin is unchanged from the flat case $z_{\rm low}({\rm min}) =
0.1$, while the maximum $z_{\rm low} ({\rm max})$ is pushed up to
higher redshifts.

\begin{figure}
\center
\epsfig{file=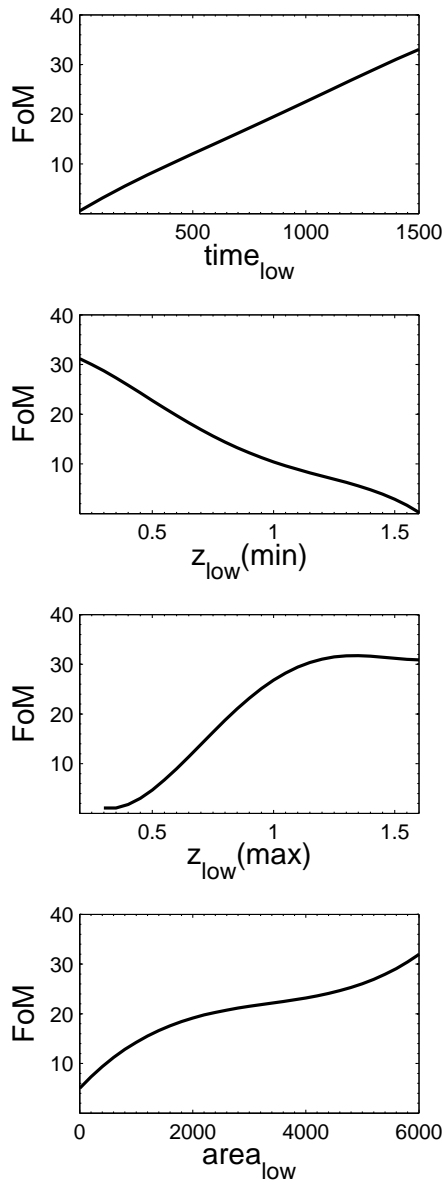, width=5.9 cm}
\caption[curved_hilo_1d]{\label{curved_hilo_1d} The
  FoM as a function of the survey parameters, where the other
  survey parameters are chose to maximize the FoM.  
  The surveys were optimized including curvature as a nuisance
  parameter.}  
\end{figure}

\subsection{The optimized survey strategy}
\label{ssec:strategy}

\begin{figure}
\center
\epsfig{file=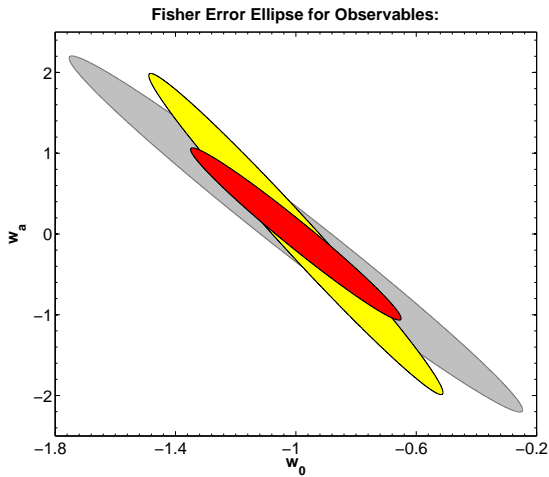, width=8 cm}
\caption[improvement]{\label{improvement} The 68\% error ellipse on
  the $w_0$ and $w_a$ parameters, with marginalization over curvature,
  for the standard WFMOS survey (grey), and the optimized
  one (red). Also shown (yellow) is the error ellipse
  were the survey optimized for a flat Universe (but the errors have
  been computed here marginalizing over curvature). The difference
between the largest ellipse and the two smaller ones shows the improvement
due to optimizing the survey for measuring the dark energy parameters,
while the difference between the smaller ellipses is due to different
cosmological models (flat or non-flat) used for the optimization. These
constraints are calculated including prior information from Planck
and SDSS.}
\end{figure}

We begin by noting the significant impact that optimization can have
in improving the science return. The FoM of the optimized case
including curvature (32, from Table~\ref{table:flatcurved}) is much larger
than the FoM of the standard baseline survey (9 from
Table~\ref{standard:table}). The amount of reduction in the area of
the error ellipse is $32/9 \simeq 3.6$, a large factor, and this is
illustrated in Figure \ref{improvement}. Put another way, the
optimized survey would reach the same dark energy equation of state 
accuracy as the standard survey after only about one-quarter of the 
survey time. The reduced accuracy of the standard survey is because 
it spends time at high redshift which would be more productively spent 
at low redshift, and its low-redshift exposures are unnecessarily long.

Having established the importance of optimization, what considerations
determine the optimal survey strategy? The principal uncertainty here
is the form of the true cosmological model. This is what we are trying
to determine, and there must clearly be competing possibilities for
the experiment to be interesting. As the optimal strategy depends on
the (unknown) true cosmological model, there will inevitably be
choices to be made which have both costs and benefits. In the context
of the models considered in this paper, the decision is whether to
assume a flat Universe or to allow for curvature; there will be a cost
if the assumption made in optimization turns out to be inappropriate
once the data are obtained and analyzed.

For the models we have considered here, the basic survey decisions are
independent of the assumed cosmological model. The first is that
high-redshift observations are unnecessary --- all survey time should
be spent at low redshift ($z<1.6$). The second is that the exposures
should be as short as possible, as this is already sufficient to
obtain the desired redshifts, and hence achieves maximal survey area.
Finally, the low-redshift limit can be taken as starting at some
suitably low value such as 0.1.

The remaining decision to be made is the upper limit of the
low-redshift bin. As we have already seen, the upper redshift limit is
different depending whether we assume flatness or
not. Table~\ref{table:flatcurved} 
gives the survey parameters for each case.

\begin{table}
\centering
\caption{Optimal survey Figure of Merit calculated in flat and curved cases, where
the optimization has been undertaken under two different assumptions, either that
$\Omega_k$ is left out or included as a nuisance parameter. The FoM in computed
including prior information from Planck and SDSS.}
\label{flatcurved:table}
\begin{tabular}{lcc} \hline
Survey optimization & without $\Omega_k$ & with $\Omega_k$ \\ \hline
FoM ($\Omega_k$ set to zero)   &  57  & 48 \\ 
FoM ($\Omega_k$ allowed to vary)   &  15   & 32 \\ \hline
\end{tabular}
\end{table}

Table \ref{flatcurved:table} shows the FoMs, now with the extra
information of the FoM that is returned if the true cosmology does not
match the assumption made in optimizing. Naturally, for a given survey
configuration, we get more accurate constraints if we assume a flat
Universe than if we allow for curvature, as the extra parameter in the
Fisher matrix dilutes the constraining power on dark energy. However,
we can now see the losses due to non-optimality. For example, if the
Universe really is flat, but we optimized to allow for curvature, our
FoM is degraded from 57 to 48. If we end up needing to allow for
curvature, having not optimized for it, the degradation is from 32 to
15 (the corresponding error ellipses for this case are shown in
Figure~\ref{improvement}).

\begin{figure}
\center
\epsfig{file=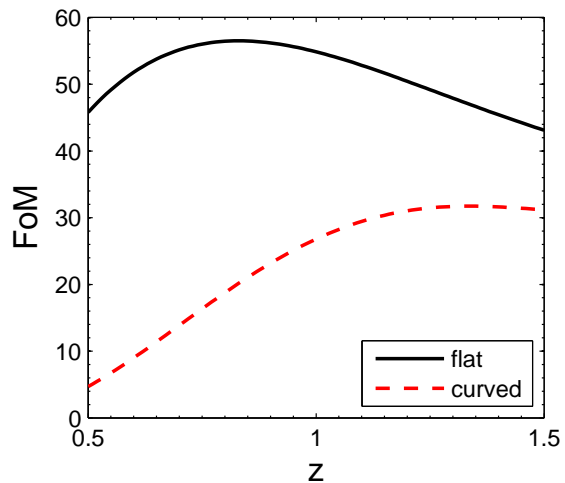, width=7.5 cm}
\caption[highest]{\label{fig:highest} The FoM as a function of the
  upper redshift limit of the survey, for both the flat case and for the
  case including curvature. All surveys use $z_{\rm min}=0.1$ and
  a minimal exposure time of 15 minutes, as discussed in the text.
Measuring the curvature requires targeting a larger redshift range.}
\end{figure}

Figure \ref{fig:highest} shows the FoMs as a function of the upper
redshift limit of the survey (reproduced from Figures \ref{flat_hilo_1d} 
and \ref{curved_hilo_1d}), 
showing the peaks at $z_{\rm max}\sim
0.7$ in the flat case and $z_{\rm max} \sim 1.35$ in the curved one.
There is no optimal way to deal with this tension, as one's opinions
as to how likely the model assumptions are governs whether the
benefits of a particular choice are likely to outweigh the costs. In
this particular case existing evidence tends to support a flat
Universe (Vardanyan, Trotta \& Silk 2009) suggesting that the
potential loss of accuracy in the flat case outweighs the ability to
measure curvature.  If we were considering different dark energy
models/parameterizations, the choice may be less clear cut.

\section{Combining WFMOS with other data sets}
\label{sec:otherdata}

Now we consider the effect of including constraints from other dark
energy surveys that will have been undertaken prior to the WFMOS-like
survey. We remind the reader that we always include SDSS and Planck 
data. Not including these data sets while allowing curvature to vary would 
change the optimal survey -- but since that data is available or will soon be, 
we would end up with a sub-optimal survey. An important question is then 
whether other planned surveys could also have a strong impact on the optimisation.

Here we consider a generic stage III type-Ia Supernovae survey
similar to that outlined in the DETF report (Albrecht et al.\ 2009), a
BAO survey similar to that expected to be completed by WiggleZ (Blake
et al.\ 2009), and another BAO survey planned to be undertaken by BOSS
(Schlegel, White \& Eisenstein 2009a). BOSS will use measurements of
the Lyman-alpha forest from quasar spectra to reconstruct the BAO at
high redshift ($z=2.5$). Since this is still somewhat speculative and
has not yet been demonstrated, we consider two cases here, one where
the QSO contribution is left out, and another where it is included.

In Section \ref{sec:results} we saw that a survey covering the range
$0.1<z<1.35$ was favoured with a high FoM. The question we ask now is
whether the predecessor experiments  will provide good enough
measurements of these low-redshift regions to drive the preferred
redshift range higher.

\begin{figure*}
\center
\epsfig{file=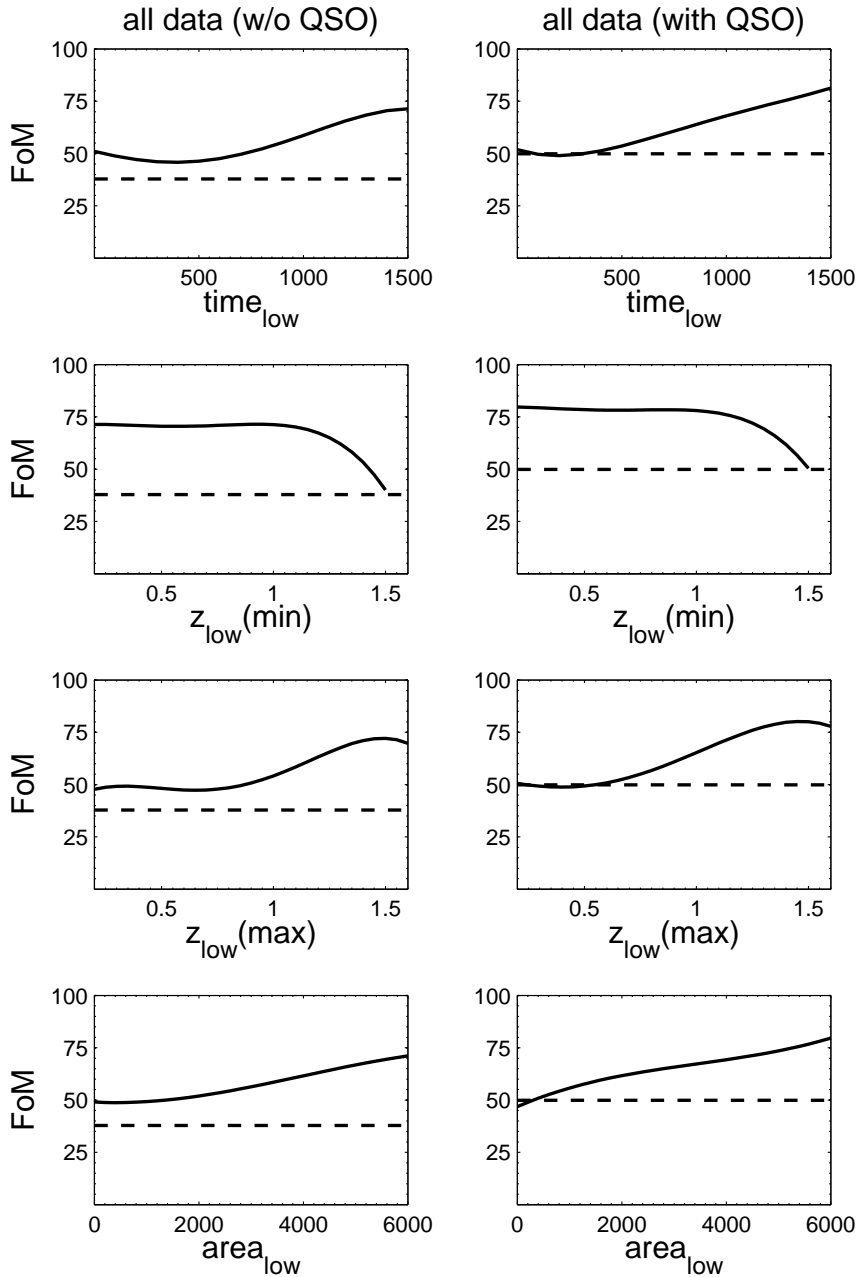, width=11.5 cm}
\caption[allnoq_withq_1d]{\label{fig:allnoq_withq_1d} The
  FoM as a function of the survey parameters when
  other experiments are included as prior information, with the QSO point 
  not included (on the left), and additionally included (on the right). The
  extra data at low redshift, $z<1$, moves the optimal WFMOS survey
  to a higher redshift of $0.9<z<1.55$.  The
  dashed lines show the FoM of the other experiments without WFMOS.}
\end{figure*}

The optimal survey parameters are shown in Table
\ref{withdata:table}. The survey parameters as a function of FoM is
shown in Figure \ref{fig:allnoq_withq_1d}, in both the cases without and
with the QSO
measurement being included
 (the full results are shown in
Figures \ref{alldata_noqso} and \ref{alldata_withqso} in Appendix~B).

\begin{table}
\centering
\caption{Best survey parameters when including other data sets. The parameters for the high redshift bin are not included as the optimal surveys spend all their time observing at low redshift. We also include the one sigma errors on the cosmological parameters predicted by the Fisher matrix approach for these optimal surveys.}
\label{withdata:table}
\begin{tabular}{lcc} \hline
Survey Parameter & SN-Ia, & SN-Ia, \\
 & WiggleZ \& & WiggleZ \& \\
 &  BOSS & BOSS (+QSO) \\ \hline
${\rm A}_{\rm low}$ (sq. degs) & 6300 & 6300 \\
$\tau_{\rm low}$ (hours) & 1500 & 1500 \\
$z_{\rm low}$ (min) & 0.9 & 0.1 \\
$z_{\rm low}$ (max) & 1.55 & 1.6 \\
exposure time (mins) & 15.0   & 15. 0 \\
number density  ($h^3$ Mpc$^{-3}$) & $6.7\times 10^{-4}$ & $6.3\times
 10^{-4}$ \\ 
number of galaxies  & $8.9 \times 10^6$ & $9.2 \times 10^6$ \\
\hline
FoM    &  72   & 80 \\ \hline 
$\sigma (w_0)$ & 0.12 & 0.21\\
$\sigma (w_a)$ & 0.41 & 0.38 \\
$\sigma (\Omega_{\rm DE})$ & 0.010 & 0.009 \\
$\sigma (\Omega_{k})$ & $1.9 \times 10^{-3}$ & $1.8\times 10^{-3}$ \\ \hline

\end{tabular}
\end{table}

Firstly we see that the maximum redshift of the low-redshift bin has increased
from $z_{\rm low} (\rm max)=1.35$ to $z_{\rm low} (\rm max) \simeq 1.55-1.6$. 
The minimum redshift of the low-redshift 
bin no longer peaks at the lowest possible value (as we see from Figure 
\ref{fig:allnoq_withq_1d}), and the FoM is
independent of its value up to $z=0.9$. Since the lower redshift range is
already covered by these other experiments, time that could have been 
spent in that range provides equal value if redeployed to $0.9 < z < 1.6$.
A high-redshift bin is
again not required, as the low-redshift bin  (at $0.9<z<1.6$)
combined with these other experiments is enough to measure the
parameters to sufficient accuracy. 

\begin{figure}
\center
\epsfig{file=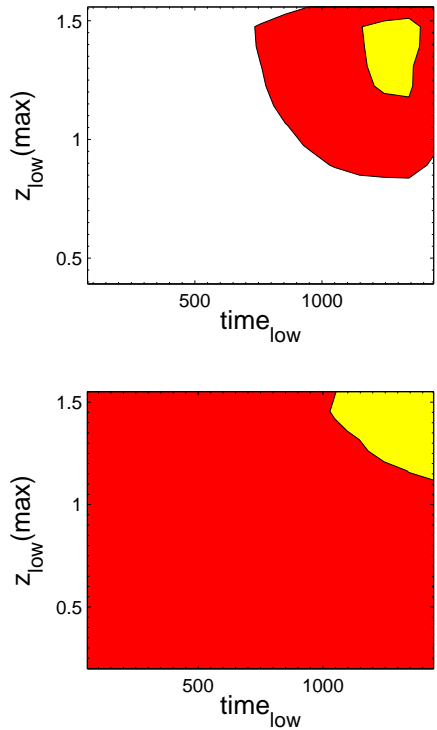, width=5.8 cm}
\caption[flexibility_plot]{\label{flexibility_plot} The 60\% and 90\%
  flexibility contours for the time and redshift for the low-redshift
  bin in the case for WFMOS by itself (upper plot), and WFMOS combined
  with other prior surveys (lower plot). We can see how the flexibility
bounds increase when the prior knowledge is stronger and the experiment
has less of an impact.}
\end{figure}

As the extra data sets are included, the flexibility bounds on the
survey parameters are expanded. Since the flexibility bounds are given
as a percentage of the peak FoM, the survey we are optimizing actually
has less of an impact on the total FoM as other data sets are
introduced (this is in contrast to parameter estimation, where more
data sets normally decrease or `tighten' the confidence limits on a
given parameter). This is very visible in Figure
\ref{flexibility_plot}, where in the case where the other data sets
are added, the 60\% flexibility bounds cover most of the possible
survey parameter space. Taken to its logical extreme, a survey which
adds little or nothing to the state of knowledge will have infinitely
large flexibility bounds, as no survey configuration will change the
FoM. Such a survey would be obsolete, and there would be little
scientific gain in undertaking it. This gives an effective `window of
opportunity' for a WFMOS-like survey, which will still make gains over
BOSS, but must be undertaken before a future all-sky dark energy
survey (SKA, LSST or JDEM/Euclid), which will have very powerful constraints
on the dark energy from a suite of observables (BAO, SN-Ia, weak
lensing and cluster number counts).

\begin{table}
\centering
\caption{FoM and parameter errors for the different surveys that
we expect to be available when the WFMOS survey takes place
(including the prior information from SDSS and Planck for each of them). 
Taking the existence of these surveys into account can significantly
change the optimization results (see text).}
\label{surveyFoM:table}
\begin{tabular}{lccc} \hline
Survey & FoM & $\sigma(w_0)$ & $\sigma(w_a)$ \\ \hline
WiggleZ & 1.1 & 1.1 & 4.25 \\
SN-Ia (stage III) & 3.7 & 0.46  & 2.5 \\ 
BOSS (no-QSO) & 6.6 & 0.47 & 1.98 \\
BOSS (+ QSO) & 21 & 0.21 & 0.71\\
WFMOS & 34 & 0.21 & 0.64 \\ \hline
Combined & 80  & 0.12 & 0.37 \\ 
\hline

\end{tabular}
\end{table}
We show the FoM and parameter errors predicted for different surveys
individually and combined in Table \ref{surveyFoM:table}.

\section{Conclusions}
\label{sec:conclusions}

We performed an analysis of the optimal dark energy survey that
could be carried out in a given time period by an experiment similar
in design to WFMOS. We estimate the accuracy of the distance 
measurements that utilize only the oscillatory part of the power spectrum 
(the 'wiggles'), using the fitting formula of Seo \& Eisenstein (2007). 
Our measure of utility, or FoM, was defined
to be proportional to the area of the error ellipse for the CPL parameters 
of the dark energy equation of state. Our results
can be summarized as follows:
\begin{itemize}
\item The high-redshift bin always gives negligible benefit, with the
  optimal surveys spending all the available time observing at low redshift,
  $z<1.6$. The $1.6<z<2$ region is the redshift desert where observations
  of star-forming galaxies are impossible, so $z=1.6$ represents a 
  hard limit on the optimization.
\item The survey area is always the maximum possible (6300 square
degrees under our assumptions) with the exposure time per field of
view always as close as possible to the minimum allowed 
(15 minutes, more than enough to obtain spectroscopic redshifts
for the majority of line-emission galaxies on an 8-metre class telescope).
\item The principal optimization decision to be made is the upper
  limit of the low-redshift range, with different values favoured
  depending on the cosmological model assumptions made.
\item Assuming a flat
universe and no external data (beyond Planck) an upper limit of $z=0.7$
is sufficient. The introduction of curvature requires the survey to push up
to $z=1.35$ for WFMOS alone.
\item The inclusion of external data sets, such as planned Supernovae
and BAO surveys whose results may predate the running of a WFMOS-like
survey, changes the optimal redshift range. The optimal maximum redshift of the
low-redshift bin is moved up to $z=1.55$. These data make
good measurements of the dark energy properties at $z<1$, the FoM
is insensitive to the minimum of the low-redshift bin as long as
$z_{\rm low} (\rm{min}) < 0.9$. 
\end{itemize}

We find some of our conclusions of the optimal BAO surveys to be 
comparable to the optimal configurations of other Dark Energy surveys.
The maximization of the survey area is the optimal configuration in both
Weak Lensing (Yamamoto et al., 2007) and ISW surveys (Douspis, et al.,  
2008). However, these types of surveys
are not so sensitive to the choice of redshift range as BAO surveys.

It would be possible to go beyond the flat $\Lambda$CDM model with the
dark energy equation of state described by something different to the
CPL parameterization. One example would be a form of $w(z)$ that
remains constant at early times before undergoing a rapid transition
at some redshift to a negative value at late times to drive the
acceleration.  This parameterization has been studied in a number of
publications (Bassett et al.\ 2002; Corasaniti \& Copeland 2003, etc).
However, the constraints on the parameters of this parameterization
are often non-Gaussian, and so the predicted constraints using a
Fisher matrix approach are often incorrect (when checked against a
more rigorous analysis using MCMC techniques). While we investigated
optimal surveys using this dark energy parameterization, the results
proved, using present methodology, to be uncomfortably unreliable.

We also showed that the flexibility bounds on the survey parameters
expand as other datasets are added in as prior information. While the
flexibility bounds should not be too narrow, as this could lead to fine
tuning of the survey which may not be realizable in practice, when the
flexibility bounds become too large it is because the instrument is
having too small an impact --- its contribution to the
total science from all surveys up to that point will be
negligible. This leaves a `window of opportunity' for a WFMOS-like
survey such that it will be of scientific benefit if it is performed
after WiggleZ and BOSS, but will become obsolete if it post-dates a
full-sky BAO survey performed by JDEM, Euclid or SKA. 

Finally, the conclusion that an optimal WFMOS-like survey should
target exclusively $0.1<z< 1.5$, aiming for the maximal possible area
and therefore the shortest possible exposure time allowing for
redshift determination, has been shown here to be quite stable. While
such a survey returns the maximal information gain on the dynamical 
dark energy parameters, other science cases could be made
for a high redshift bin, such as using redshift space distortions to
probe the growth of structure, and so the theory of gravity. The 
Dark Energy optimal survey 
would also not be as good for other science goals like galaxy evolution, 
which desire high signal-to-noise
spectra rather than redshifts alone. As we have shown that such deeper
exposures are of negligible benefit to the dark energy FoM, an
instrument aiming to carry out both types of science would need to do
so via separate survey programmes, rather than by sharing of a single
dataset.

\section*{Acknowledgments}

This research was undertaken as part of the Conceptional Design Phase 
(CoDR) for the Gemini/Subaru WFMOS instrument and was partly 
funded via AURA Contract No. 0525280-GEM00467. DP received
support under this contract. We are grateful to the WFMOS ``team A" for their help 
and guidance during this CoDR.
We acknowledge Andrey Kaliazin for debugging
assistance.  DP, MK, ARL and RCN are supported by
STFC. MV is supported by the Raffy Manoukian Scholarship and partially 
supported by the Philip Wetton Scholarship at Christ Church, Oxford.
The analysis was performed on the Glamdring cluster of Oxford
University, the Archimedes computing cluster
at the University of Sussex, supported by funds from SRIF3, and the
COSMOS supercomputer in Cambridge, supported by SGI, Intel, HEFCE and
STFC.


\appendix

\section{Fisher matrix formalism}
\label{Fisher:appendix}

As in our previous paper (P07), we use a Fisher matrix approach to
compute the predicted experimental constraints on the cosmological
parameters. The Fisher matrix is defined to be  
\beq
F_{AB} = \frac{\dd^2(-\ln\LL)}{\dd\theta_A\dd\theta_B} \,,
\eeq
where $\LL$ is the likelihood, and $\theta_i$ is one of the
cosmological parameters. By the Cramer--Rao bound, the inverse of the
Fisher matrix gives an estimator of the smallest (co)variance of the
parameters. 

In our previous work we computed the Fisher matrix exactly, but
considerations of the parameter set being used and the experimental
data that could be included meant that in this work we compute the
Fisher matrix elements numerically.\footnote{Similar work by Bassett
et al.\  (2009) has made such numerical computation available as a
MATLAB toolbox (called Fisher4Cast), available at
http://www.cosmology.org.za/} 

We do this by first gathering up all the experimental data that will
be considered (the WFMOS survey configuration under consideration,
Planck, the SDSS BAO point, plus whatever other datasets we are
including), and simulating their observables ($d_{\rm A}$ \&
$H(z)$ for BAO experiments, $R$ \& $l_a$ for Planck, $d_l$ for
supernovae experiments) at the fiducial cosmology, with no scatter in
the mean values. Since the Fisher matrix is defined as the expectation
of the Hessian of the log-likelihood, averaged over all possible
realizations of the data, this averaging process removes the scatter
in the mean of the data point. Then we use this to compute the
likelihood in the region around the fiducial cosmology.

We can compute the slope of a function through a finite-difference
operation. For example, the slope of a function $f$ evaluated at $x$
can be given by 
\beq
f'(x) = \frac{f(x+\epsilon)-f(x-\epsilon)}{2\epsilon} \,,
\eeq
where $\epsilon$ is some small positive number. As we are
evaluating the likelihood at the fiducial cosmology, which should be
identical to the maximum likelihood point, the slope of the likelihood
in any direction should be zero, or as close to it as numerical
accuracy will allow. Here we use the finite-difference method to
compute the second derivative of the likelihood, evaluated at the
fiducial cosmology. 

For each individual Fisher matrix element we take steps of size
$\epsilon$ in both cosmological parameters. We can estimate the slope
in the direction of one of the parameters at the displaced point of
the other, i.e. 
\begin{eqnarray}
\frac{\dd(-\ln\LL(\theta_B \pm \epsilon))}{\dd\theta_A} = \hspace*{4.5cm}\\
\quad -\frac{\ln\LL(\theta_B \pm \epsilon, \theta_A + \epsilon)
  -\ln\LL(\theta_B \pm \epsilon, \theta_A - \epsilon)}{2\epsilon}
\,. \nonumber
\end{eqnarray}
(This holds also if $\theta_A$ and $\theta_B$ are exchanged.)
 From here our estimate of the second derivative is simply the finite
difference of the slopes, evaluated at the slightly displaced
positions, 
\beq
\frac{\dd^2(-\ln\LL)}{\dd\theta_A\dd\theta_B} = \frac{1}{2\epsilon}
\left[ 
\frac{\dd(-\ln\LL(\theta_B + \epsilon))}{\dd\theta_A}
-\frac{\dd(-\ln\LL(\theta_B - \epsilon))}{\dd\theta_A}\right]. 
\eeq
We tune the step parameters $\epsilon$ to be small enough to achieve
numerical convergence of the Fisher matrix. 

\section{Full presentation of results}
\label{Plots:appendix}

\begin{figure*}
\center
\epsfig{file=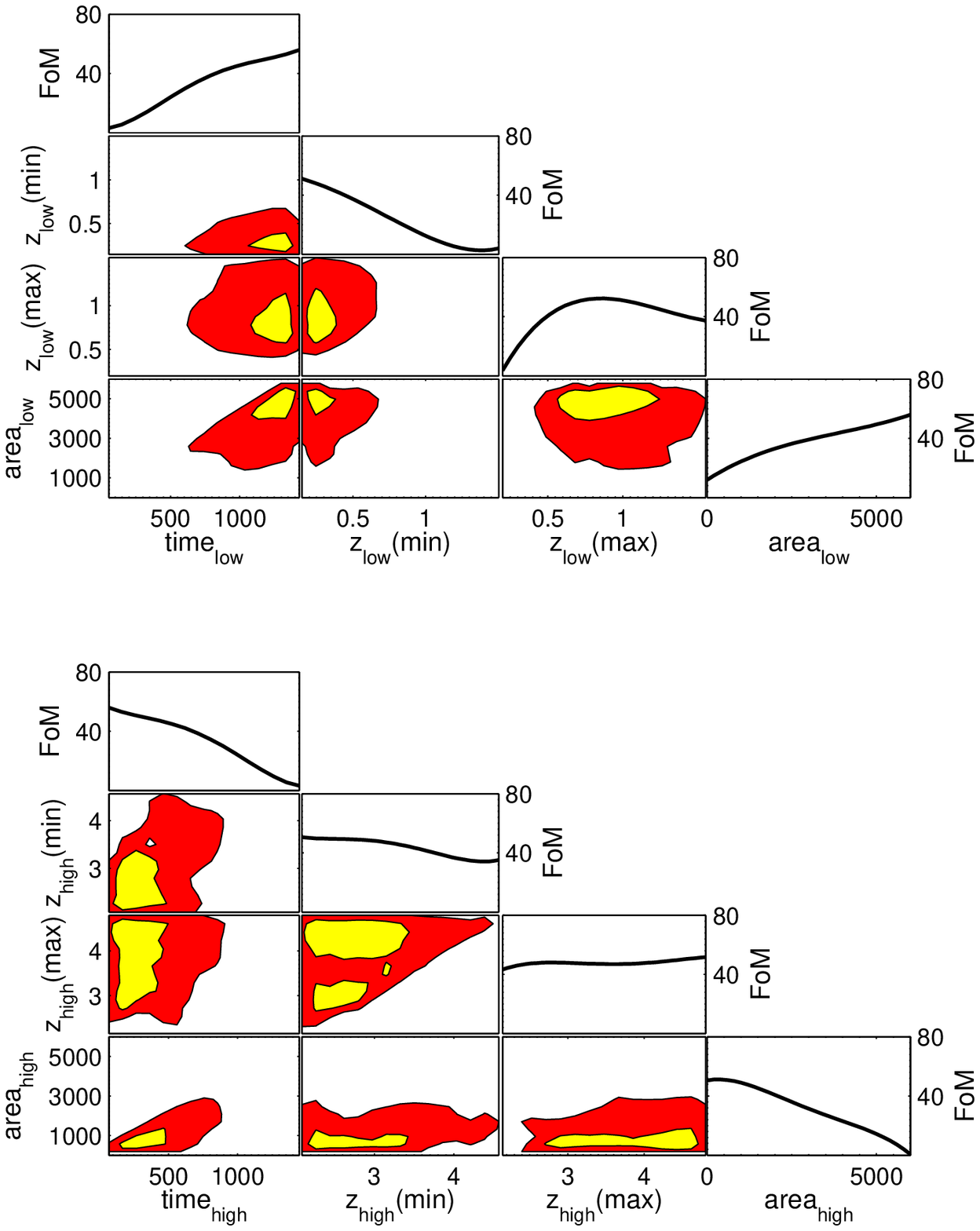, width=16 cm}
\caption[flat_hilo]{\label{flat_hilo} The optimal FoM as a
  function of the survey parameters (where the other parameters are chosen
  to maximize the FoM), for surveys optimized
  for a flat universe. The 2-d contours delimit the 90\% and 60\% flexibility bounds.
  We see the optimal survey is one that spends all its time at low redshift (${\rm time}_{\rm low} = 1500 {\rm hrs}$) and 
  maximizes the area in the low-redshift bin. We see that the high-redshift bin adds nothing to the FoM,
  and so it is insensitive to those parameters, except the time in the high-redshift bin which is
  minimized.} 
\end{figure*}

\begin{figure*}
\center
\epsfig{file=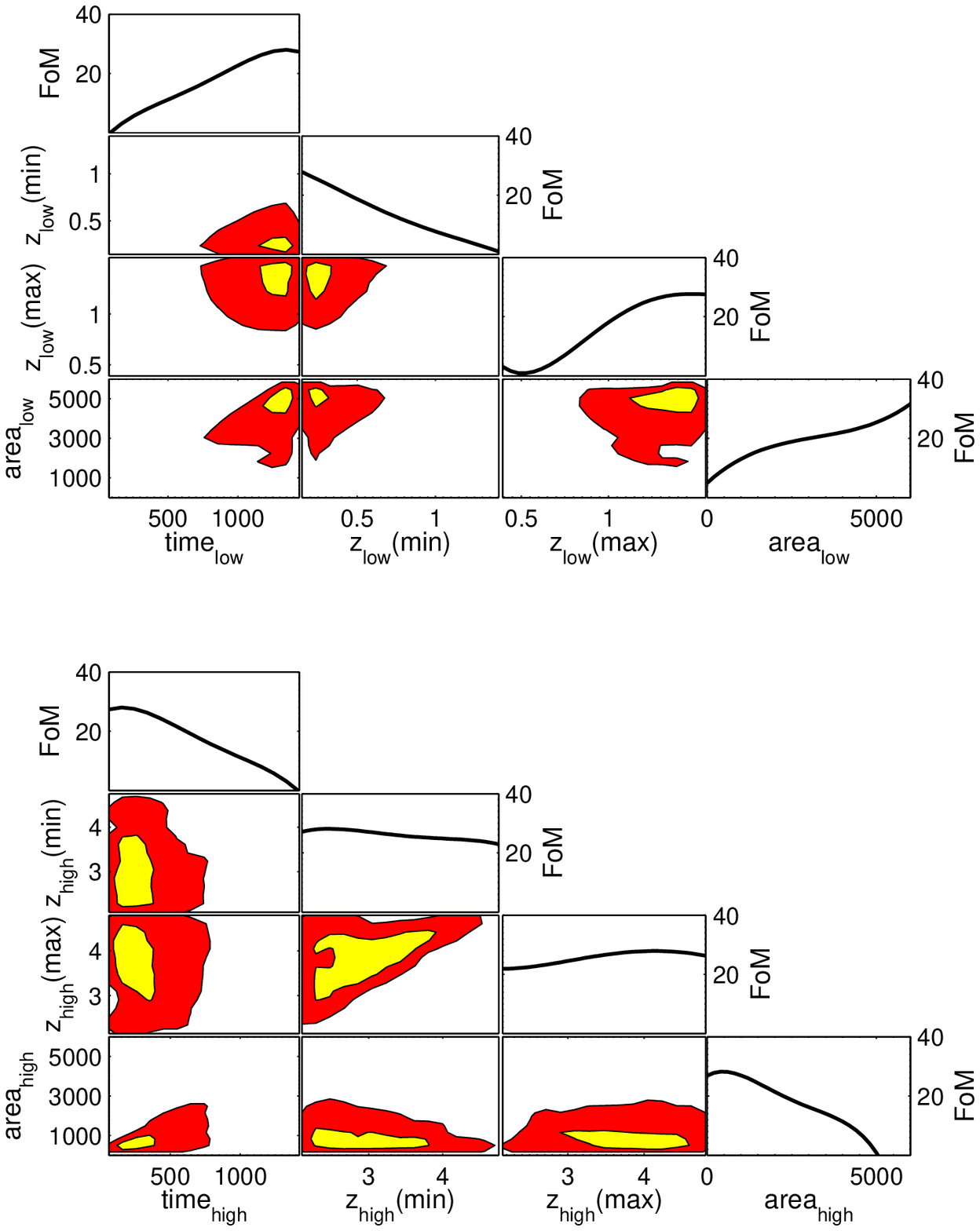, width=16 cm}
\caption[curved_hilo]{\label{curved_hilo} The FoM as a
  function of the survey parameters (where the other parameters are chosen
  to maximize the FoM), for surveys optimized
  including curvature as a nuisance parameter. The 2-d contours delimit the 90\% and 60\% flexibility bounds.
  We see the optimal survey is one that spends all its time at low redshift and 
  maximizes the area in the low-redshift bin, but now the optimal maximum of the low-redshift bin is moved up to $z=1.35$. As before, we see that the high-redshift bin adds nothing to the FoM,
  and so it is insensitive to those parameters, except the time in the high-redshift bin which is
  minimized.} 
\end{figure*}

\begin{figure*}
\center
\epsfig{file=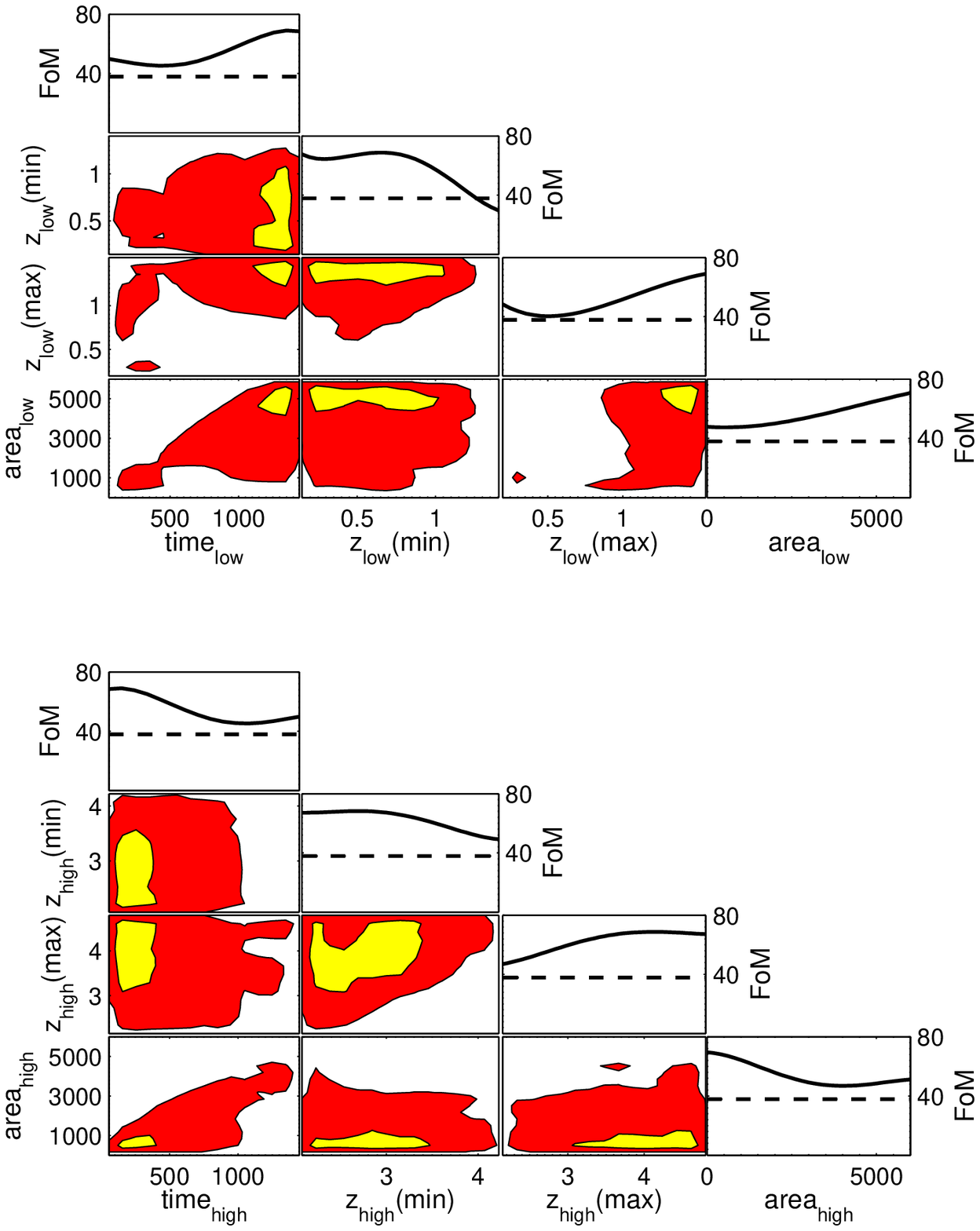, width=16cm}
\caption[alldata_nosqo]{\label{alldata_noqso} The FoM as a 
  function of the survey parameters for a WFMOS-like survey with other data
  sets (SN-Ia, BOSS, but without the QSO point at z=2.5, and WiggleZ)
  included as external datasets. The dashed line shows the FoM of the
  other surveys combined (including Planck), but without WFMOS. 
  The 2-d contours delimit the 90\% and 60\% flexibility bounds.
  We see the optimal survey is one that spends all its time at low redshift and 
  maximizes the area in the low-redshift bin. The FoM now peaks at  $z=1.55$ for
  the maximum of the low-redshift bin, but is insensitive to the minimum as long as  $z_{\rm low} (\rm{min})\le0.9$. 
  As before, we see that the high-redshift bin adds nothing to the FoM,
  and so it is insensitive to those parameters, except the time in the high-redshift bin which is
  minimized.} 
\end{figure*}

\begin{figure*}
\center
\epsfig{file=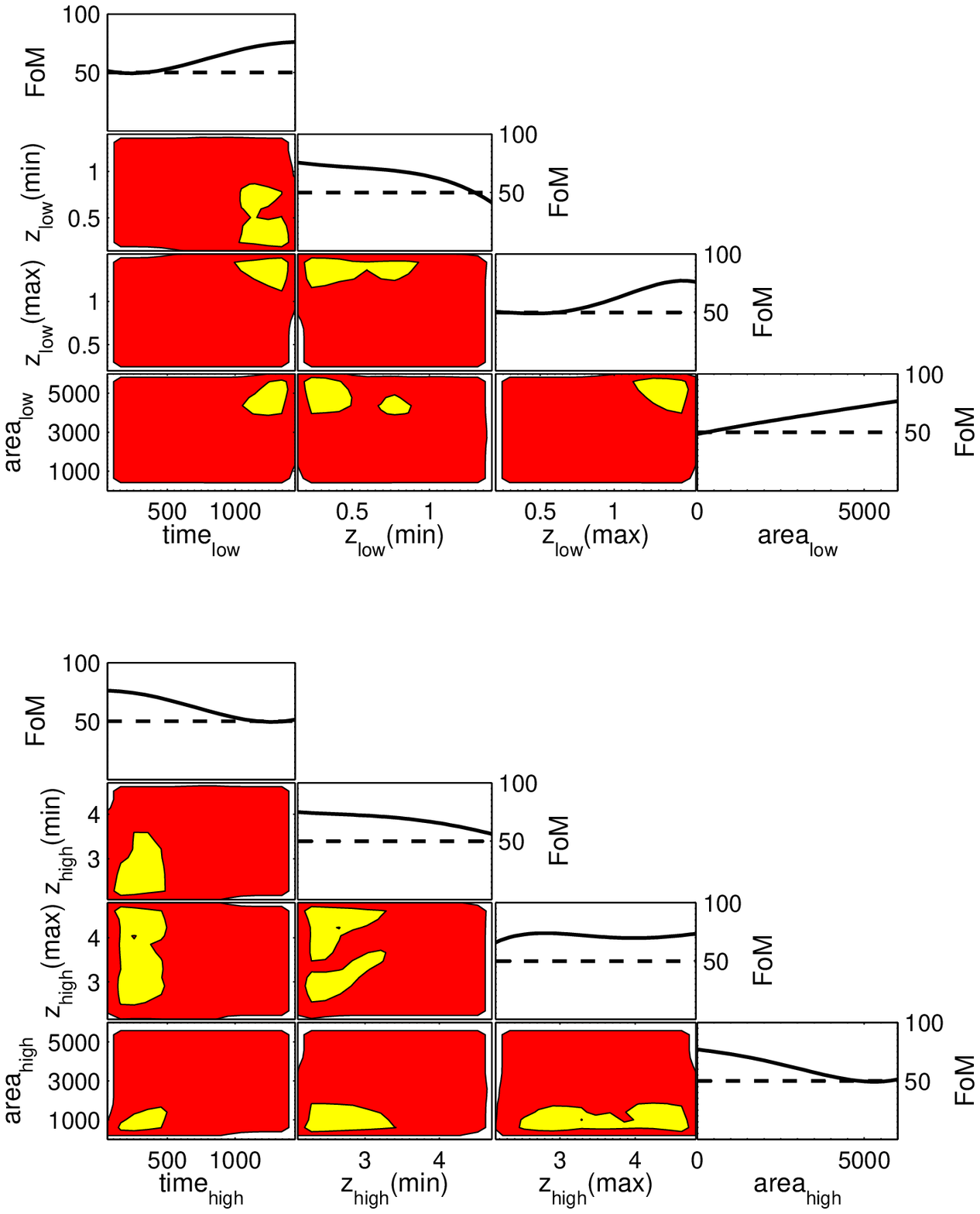, width=16cm}
\caption[alldata_withqso]{\label{alldata_withqso} The FoM
  as a function of the survey parameters for a WFMOS-like survey with other
  data sets (SN-Ia, BOSS with the QSO point at z=2.5 and WiggleZ)
  included as external datasets. The dashed line shows the FoM of the
  other surveys combined (including Planck), but without WFMOS. 
   The 2-d contours delimit the 90\% and 60\% flexibility bounds.
  We see the optimal survey is one that spends all its time at low redshift and 
  maximizes the area in the low-redshift bin. The FoM now peaks at  $z=1.6$ for
  the maximum of the low-redshift bin, but is insensitive to the minimum 
  as long as  $z_{\rm low} (\rm{min})\le0.9$. Notice
  that the 60\% flexibility bound now covers most of the survey
  parameter space.}
\end{figure*}

\end{document}